\documentclass[eqsecnum,aps,pra,twocolumn,groupedaddress,showpacs,showkeys]
{revtex4}

\usepackage{graphicx}
\begin{document}

\newcommand{\be}{\begin{equation}}
\newcommand{\ee}{\end{equation}}
\def\cstok#1{\leavevmode\thinspace\hbox{\vrule\vtop{\vbox{\hrule\kern1pt
\hbox{\vphantom{\tt/}\thinspace{\tt#1}\thinspace}}
\kern1pt\hrule}\vrule}\thinspace}

\title{From Rindler space to the electromagnetic energy-momentum
tensor of a Casimir apparatus in a weak gravitational field}

\author{Giuseppe Bimonte$^{1,2}$ \thanks{
Electronic address: giuseppe.bimonte@na.infn.it},
Giampiero Esposito$^{2}$ \thanks{
Electronic address: giampiero.esposito@na.infn.it},
and Luigi Rosa$^{1,2}$ \thanks{
Electronic address: luigi.rosa@na.infn.it}}

\affiliation{${ }^{1}$Dipartimento di Scienze Fisiche, Universit\`{a} di
Napoli Federico II, Complesso Universitario
di Monte S. Angelo, Via Cintia, Edificio 6,
80126 Napoli, Italy\\
${ }^{2}$INFN, Sezione di Napoli, Complesso Universitario di Monte
S. Angelo, Via Cintia, Edificio 6, 80126 Napoli, Italy}

\date{\today}

\begin{abstract}
This paper studies two perfectly conducting parallel plates in the weak
gravitational field on the surface of the Earth. Since the appropriate
line element, to first order in the constant gravity acceleration g,
is precisely of the Rindler type, we can exploit the formalism for studying
Feynman Green functions in Rindler spacetime. Our analysis does not reduce
the electromagnetic potential to the transverse part before quantization.
It is instead fully covariant and well suited for obtaining all components
of the regularized and renormalized energy-momentum tensor to arbitrary
order in the gravity acceleration g. The general structure of the
calculation is therefore elucidated, and the components of the Maxwell
energy-momentum tensor are evaluated up to second order in g, improving
a previous analysis by the authors and correcting their old first-order
formula for the Casimir energy.
\end{abstract}



\maketitle

\section{Introduction}

Although quantum field theory in curved spacetime is an hybrid framework,
being based on the coupling of the classical Einstein tensor to a quantum
concept like the vacuum expectation value of the regularized and
renormalized energy-momentum tensor, it has led to exciting developments
over many years \cite{Fulling1989}, \cite{GR-QC-0610008}, \cite{08032003}.
In particular, the theoretical
discovery by Hawking of particle creation by black holes
\cite{CMPHA-43-199} is a peculiar phenomenon of quantum field theory in
curved spacetime, and all modern theories of quantum gravity face the
task of evaluating and understanding black hole entropy and the ultimate
fate of black holes.

Since the chief goal of quantum field theory in curved spacetime may be
regarded as being the evaluation of the energy-momentum tensor
\cite{PRPLC-19-295} on the right-hand side of the semiclassical Einstein
equations
\begin{equation}
R_{\mu \nu}-{1\over 2}g_{\mu \nu}R=8 \pi G \langle T_{\mu \nu} \rangle,
\label{(1.1)}
\end{equation}
it is very important even nowadays to look at problems where new physics
(at least in principle) can be learned or tested while using Eq. (1.1).
In particular, we are here concerned with a problem actively investigated
over the last few years, i.e. the behavior of rigid Casimir cavities
in a weak gravitational field \cite{ASTRO-PH-0209312},
\cite{IMPAE-A17-804}, \cite{PHLTA-A297-328},
\cite{CQGRD-22-5109}, \cite{PHRVA-D74-085011}, \cite{PHRVA-D76-025004},
\cite{PHRVA-D76-025008}, \cite{JPAGB-40-10935}. An intriguing theoretical
prediction is then found to emerge, according to which Casimir energy
obeys exactly the equivalence principle \cite{PHRVA-D76-025004},
\cite{PHRVA-D76-025008}, \cite{JPAGB-40-10935}, and the Casimir apparatus
should experience a tiny push (rather than being attracted) in the upwards
direction. The formula for the push has been obtained in three different
ways, i.e.,
\vskip 0.3cm
\noindent
(i) A heuristic summation over modes \cite{IMPAE-A17-804};
\vskip 0.3cm
\noindent
(ii) A variational approach \cite{PHRVA-D76-025004};
\vskip 0.3cm
\noindent
(iii) An energy-momentum analysis \cite{PHRVA-D74-085011}.

While all approaches now agree about the push and the magnitude
of the effect \cite{PHRVA-D76-025008}, the work in Ref.
\cite{PHRVA-D74-085011}, despite its explicit analytic formulas for
$\langle T_{\mu \nu} \rangle$, led to legitimate puzzlements, being
accompanied by a theoretical prediction of nonvanishing trace anomaly.
It has been therefore our aim to perform a more careful investigation
of the energy-momentum tensor of our rigid Casimir apparatus. The
nonvanishing trace will be shown to result from a calculational mistake,
and a better understanding of the general calculation, possibly to all
orders in g, will be gained. Sections II, III and IV describe basic
material on Rindler coordinates, scalar and photon Green functions.
Ward identities are checked explicitly in Sec. V, while the various
parts of the energy-momentum tensor are analyzed in Secs. VI and VII.
Concluding remarks are presented in Sec. VIII, and relevant details
are given in the appendices.

\section{Rindler coordinates}

We work in natural units, in which $\hbar=c=1$. In these units,
the gravity acceleration has dimensions of an inverse length.
Neglecting tidal forces, the weak gravitational field on the
surface of the Earth is described by the line element
\be
ds^2=-
\left(1+ {2{\rm g}\,z} \right)  d t^2 +d z^2+d {\bf
x}_{\perp}^2\;,
\label{earth}
\ee
where ${\rm g}$ is the gravity
acceleration, and ${\bf x}_{\perp} \equiv (x,y)$. We consider an ideal
Casimir apparatus, consisting of two perfectly reflecting mirrors
lying in the horizontal plane, and separated by an empty gap of
width $a$. We let the origin of the $z$ coordinate coincide with
the lower mirror, in such a way that the mirrors have coordinates
$z=0$ and $z=a$, respectively. To first order in the small
quantity ${\rm g} z $, the line element in Eq. (2.1)
coincides with the Rindler metric
\be
ds^2=- \left(\frac{\xi}{\xi_1}\right)^2 d t^2 +d \xi^2+d
{\bf x}_{\perp}^2\;,\label{rindler}
\ee
where
\be
\xi \equiv \frac{1}{{\rm g}}+z\equiv \xi_1+z.
\label{change}
\ee
In the Rindler coordinates, the plates are located at
\be
\xi_1 \equiv \frac{1}{{\rm g}},\;\;\;\;\xi_2 \equiv \xi_1+a.
\ee
The time coordinate $t$ in
Eq. (2.2) therefore represents the proper time for an
observer comoving with the mirror at $\xi_1$. In what follows, it
shall be often convenient to work out exact formulas for a Casimir
apparatus in the Rindler gravitational field, and to recover the
corresponding formulas for the weak field in Eq. (2.1) by
taking the large $\xi_1$ limit of the Rindler results. In the rest
of the paper we shall use the following notations:
$ \mu,\nu\;\;{\rm range\;over}\;(t,\xi,x,y)$,
$ a,b\;\;{\rm range\;over}\;(t,\xi)$,
$ i,j\;\;{\rm range\;over}\;(x,y).$

\section{Scalar Green functions}

We consider the Green functions $G^{(D)}(x,x')$ and
$G^{(N)}(x,x')$ for a massless scalar field propagating in the Rindler
metric and satisfying Dirichlet and Neumann
boundary conditions, respectively, i.e.
\be
\cstok{\ }\,G^{(D/N)}(x,x')
=-g^{-1/2}\,\delta(x,x'),\label{eom}
\ee
\be
G^{(D)}(x,x')\vert_{z=z_i}=0\;\;\;\;i=1,2
\ee
\be
\partial_{z}G^{(N)}(x,x')\vert_{z=z_i}= 0\;\;\;\;i=1,2
\ee
By virtue of translation invariance in the $t,x,y$ directions,
they can be written as
\begin{widetext}
\be
G^{(D/N)}(x,x')= {\xi_1} \int \frac{d \omega}{2 \pi} \exp[-i
\omega(t-t')] \int \frac{d^2{\bf k}}{(2 \pi)^2} \exp[i {\bf
k}\cdot ({\bf x}_{\perp}-{\bf x'}_{\perp})]\,
\chi^{(D/N)}(\xi,\xi'|i\nu,k),
\ee
\end{widetext}
where ${\bf k} \equiv (k_x,k_y)$, $k \equiv \sqrt{k_x^2+k_y^2}$
and $\nu \equiv \xi_1\,
\omega $. The functions $\chi^{(D/N)} (\xi,\xi'|i\nu,k)$ satisfy
the equation
\be
\left[\xi \frac{d}{d \xi}\left(\xi\frac{d}{d
\xi}\right)+(\nu^2-\xi^2 k^2)\right]{\chi}^{(D/N)} (\xi,\xi'|i
\nu,k)=-\xi\,\delta(\xi-\xi'),
\label{bessel}
\ee
together with the boundary conditions
\be
{\chi}^{(D)} (\xi_i,\xi'|i
\nu,k)=\frac{d}{d \xi}{\chi}^{(N)} (\xi_i,\xi'|i \nu,k)=0.
\ee
The functions $\chi^{(D/N)} (\xi,\xi'|i \nu,k)$
can be expressed in
terms of the modified Bessel functions of imaginary order $I_{i
\nu}(k \xi)$ and $K_{i \nu}(k \xi)$ that are solutions of the
homogeneous equation corresponding to Eq. (\ref{bessel}). We
define the function
\be
W_{i \nu}(u,v) \equiv K_{i
\nu}(u)I_{i \nu}(v)-I_{i \nu}(u)K_{i \nu}(v)\,.
\ee
Thus we have
\be
{\chi}^{(D)} (\xi,\xi'|i \nu,k)=-\frac{W_{i\nu}(k \xi_>,k
\xi_2)W_{i\nu}(k \xi_<,k \xi_1)}{W_{i\nu}(k \xi_1,k
\xi_2)},
\label{dirbr}
\ee
\be
{\chi}^{(N)} (\xi,\xi'|i\nu,k)
=-\frac{(\partial_v W_{i\nu})(k \xi_>,k \xi_2)(\partial_v
W_{i\nu})(k \xi_<,k \xi_1)}{(\partial_u
\partial_v W_{i \nu})(k \xi_1,k \xi_2)},
\label{neubr}
\ee
where $\xi_{>} \equiv {\rm max}(\xi,\xi')$
and $\xi_{<} \equiv {\rm min} (\xi,\xi')$.

By using the identities
\be
K_{i \nu}(e^{i \pi} \zeta)=e^{\nu
\pi}K_{i \nu}(\zeta)-i\pi I_{i \nu}(\zeta),\ee \be K'_{i \nu}(e^{i
\pi} \zeta)=-e^{\nu \pi}K'_{i \nu}(\zeta)
+i\pi I'_{i\nu}(\zeta),
\ee
with $'$ denoting differentiation, to eliminate
$I_{i \nu}$ and $I_{i \nu}'$ from Eqs. (\ref{dirbr}) and
(\ref{neubr})), the propagators can be expressed in the following form:
\be
G^{(D/N)}(x,x')=G^{(0)}(x,x')+\tilde{G}^{(D/N)}(x,x')\;,
\label{split}
\ee
where $G^{(0)}(x,x')$ is the Feynman propagator for a massless
scalar field in Minkowski space time \cite{JMAPA-17-2101},
\cite{PRSLA-A354-79}
\begin{widetext}
\be
G^{(0)}(x,x')=\frac{i \xi_1}{\pi}\int \frac{d \omega}{2 \pi}
\exp[-i \omega(t-t')] \int \frac{d^2{\bf k}}{(2 \pi)^2} \exp[i
{\bf k}\cdot ({\bf x}_{\perp}-{\bf x'}_{\perp})]K_{i \nu}(k
\xi_>)K_{i \nu}(e^{i \pi}k \xi_<),
\ee
\end{widetext}
and
\begin{widetext}
\be
\tilde{G}^{(D/N)}(x,x')= {\,\xi_1} \int \frac{d \omega}{2 \pi}
\exp[-i \omega(t-t')] \int \frac{d^2{\bf k}}{(2 \pi)^2} \exp[i
{\bf k}\cdot ({\bf x}_{\perp}-{\bf
x'}_{\perp})]\tilde{\chi}^{(D/N)} (\xi,\xi'|i \nu,k) ,
\label{tildeG}
\ee
\end{widetext}
where
\be
\tilde{\chi}^{(D)} \!=  \frac{i}{\pi}\frac{{\cal
A}^{(D)} (\xi,\xi'|i \nu,k)}{K_{i \nu}(e^{i\pi} k\xi_1)K_{i
\nu}(k\xi_2)-K_{i \nu}(k\xi_1)K_{i \nu}(e^{i\pi}
k\xi_2)}
\label{tildechiD}
\ee
having set
$$
{\cal A}^{(D)} (\xi,\xi'|i \nu,k) \equiv K_{i \nu}(k\xi)K_{i
\nu}(e^{i\pi} k\xi_1)[K_{i \nu}(k\xi')K_{i \nu}(e^{i\pi} k\xi_2)$$
$$-K_{i \nu}( e^{i\pi}k\xi')K_{i \nu}( k\xi_2)]+K_{i
\nu}(e^{i\pi}k\xi)K_{i \nu}( k\xi_2)
$$
\be
\times[K_{i \nu}(e^{i\pi}k\xi')K_{i
\nu}(k\xi_1)-K_{i\nu}(k\xi')K_{i
\nu}(e^{i\pi}k\xi_1)],
\label{AD}
\ee
and
\be
\tilde{\chi}^{(N)} =
\frac{i}{\pi}\frac{{\cal A}^{(N)} (\xi,\xi'|i \nu,k)}{K'_{i\nu}(
k\xi_1)K'_{i\nu}(e^{i\pi}k\xi_2)-K'_{i\nu}(e^{i
\pi}k\xi_1)K'_{i\nu}( k\xi_2)}
\label{tildechiN}
\ee
where
$$
{\cal A}^{(N)} (\xi,\xi'|i \nu,k) \equiv K_{i\nu}(k\xi)K'_{i\nu}(e^{i\pi}
k\xi_1)[K_{i\nu}(k\xi')K'_{i\nu}(e^{i\pi} k\xi_2)$$ $$+K_{i\nu}(
e^{i\pi}k\xi')K'_{i\nu}( k\xi_2)]
+K_{i\nu}(e^{i\pi}k\xi)K'_{i\nu}(k\xi_2)
$$
\be
\times[K_{i\nu}(e^{i\pi}k\xi')K'_{i\nu}(k\xi_1)
+K_{i\nu}(k\xi')K'_{i\nu}(e^{i\pi}k\xi_1)].
\label{AN}
\ee
As is clear from Eqs. (\ref{AD}) and (\ref{AN}),  the
quantities $\tilde{\chi}^{(D/N)} (\xi,\xi')$ are symmetric
functions of $\xi$ and $\xi'$, and are both regular at $\xi=\xi'$.
The integrands for $\tilde{G}^{(D/N)}(x,x')$ in Eq. (\ref{tildeG})
have simple poles at the zeros of the quantities that occur in the
denominators of the expressions for $\tilde{\chi}^{(D/N)} $,
Eqs. (\ref{tildechiD}) and (\ref{tildechiN}). These zeros are all
located on the real $\nu$ axis. The Feynman propagator is obtained
by deforming the contour for the $\nu$-integration, in such a way
that it passes below the poles on the negative $\nu$-axis and
above those on the positive $\nu$-axis.

\section{The photon propagator}

We quantize the classical solutions of the field equations \be
\nabla_{\mu} \nabla^{\mu} A_{\nu}(x)=0\;,\;\;\;\xi_1 \le \xi \le
\xi_2 \ee which are obtained on choosing the Lorenz gauge
\cite{PHMAA-34-287}, subject to the boundary conditions \be
A_{\tau}(\xi_i)=A_j({\xi_i})=0\,,\;\;\;\partial_{\xi}(\xi
A_{\xi})(\xi_i)=0 \label{bc}. \ee Equation (\ref{bc}) expresses
the mixed boundary conditions on the potential corresponding to
the choice of perfect conductor boundary conditions
\cite{FTPHD-85-1}. They are preserved under gauge transformations
provided that the Faddeev--Popov ghost fields $\chi$ and $\psi$
obey homogeneous Dirichlet conditions on the plates
\cite{FTPHD-85-1}. The modes are normalized via the following
Klein--Gordon inner product: \be (w,v) \equiv i \int d^2{\bf x}
\int_{\xi_1}^{\xi_2}d \xi\, \frac{ \xi_1}{\xi}\, w^{\mu*}
\stackrel{\leftrightarrow}{\nabla}_{t}v_{\mu}\;. \label{inner} \ee
Note that the above inner product is not positive definite, by
virtue of the Lorentz signature of the metric. A convenient basis
of gauge fields $A_{\mu}$ can be obtained in terms of the
normalized modes for the Dirichlet and Neumann scalar problems,
$\phi^{(D)}_{r{\bf k}}(x)$ and $\phi^{(N)}_{r{\bf k}}(x)$
respectively: \be \phi^{(D/N)}_{r{\bf
k}}(x)=\exp[-i\,\omega^{(D/N)}_{rk}t+i\, {\bf k}\cdot{\bf
x}_{\perp}]\,\tilde{\phi}^{(D/N)}_{r{\bf k}}(\xi). \ee These modes
obey the differential equation (cf. Eq. (3.5)) \be
\left[\frac{\xi}{\xi_1} \frac{d}{d
\xi}\left(\frac{\xi}{\xi_1}\frac{d}{d \xi}\right)+\left(
{\omega^{(D/N)}_{rk}} \right)^2 -\left(\frac{\xi}{\xi_1}\right)^2
k^2\right]\tilde{\phi}^{(D/N)}_{r{\bf {\bf k}}}(\xi)=0, \ee the
boundary conditions \be \tilde{\phi}^{(D)}_{r{\bf
k}}(\xi_i)=\frac{d}{d \xi}\tilde{\phi}^{(N)}_{r{\bf k}}(\xi_i)=0,
\ee and the orthogonality relation
$$  \int d^2{\bf x}
\int_{\xi_1}^{\xi_2} d \xi\,\frac{
\xi_1}{\xi}\,\phi^{(D/N)*}_{r{\bf k}}(x) \phi^{(D/N)}_{r'{\bf k}'}(x)
$$
\be
=\frac{1}{2\,\omega^{(D/N)}_{rk}}\, \delta_{r r'} (2
\pi)^2 \delta({\bf k}-{\bf k}').
\ee
We obtain
\be
A_{\mu}=\sum_{r=1}^{\infty} \int \frac{d^2 {\bf k}}{k(2
\pi)^2}\sum_{\lambda=0}^3[{ A}_{r{\bf
k}\mu}^{(\lambda)}(x)a_{r\lambda}({\bf k})+{A}_{r{\bf k}
\mu}^{(\lambda)*}(x)a_{r\lambda}^*({\bf k})],
\ee
where
\be
{A}_{r{\bf k} \mu}^{(0)}(x) =(\nabla_a,0)\,
\phi^D_{r{\bf k}}(x),
\ee
\be { A}_{r{\bf k}\mu}^{(1)}(x)
=(p_a,0)\, \phi^N_{r{\bf k}}(x),
\ee
\be
{A}_{r{\bf k} \mu}^{(2)}(x)
=(0,p_i)\, \phi^D_{r{\bf k}}(x),
\ee
\be
{A}_{r{\bf k} \mu}^{(3)}(x) =(0,\nabla_i)\,
\phi^D_{r{\bf k}}(x),
\ee
where $p_a=\epsilon_{ab} \nabla^b$, $p_i=\epsilon_{ij}
\nabla^j$, with
\be
\epsilon_{ab} \equiv \frac{1}{\xi_1}\left(\begin{array}{cc}
0 & \xi \\
-\xi & 0 \\
\end{array}\right)
\ee
and
\be
\epsilon_{ij} \equiv \left(\begin{array}{cc}
0 & 1 \\
-1 & 0 \\
\end{array}\right).
\ee
The above modes satisfy the orthogonality relations
\be
({ A}_{r{\bf k}}^{(\lambda)}, { A}_{r'{\bf
k}'}^{(\lambda')})=\eta^{\lambda \lambda'} \delta_{rr'} (2 \pi)^2
\delta({\bf k}-{\bf k}')\,,
\ee
\be
({ A}_{r{\bf k}}^{(\lambda)}, {
A}_{r'{\bf k}'}^{(\lambda ')*})= 0\,,
\ee
where $\eta^{\lambda
\lambda'}=\eta_{\lambda \lambda'}={\rm diag}(-1,1,1,1)$.

It should be stressed that, despite some formal analogies with the
work in Ref. \cite{PRSLA-A354-79}, we are not reducing the theory to
its physical degrees of freedom before quantization.
In quantum theory the amplitudes $a_{r \lambda}({\bf k})$ are
replaced by operators satisfying the commutation relations
\be
[a_{r\lambda}({\bf k}),a^*_{r'\lambda'}({\bf
k'})]=\eta_{\lambda \lambda'}\delta_{rr'} (2 \pi)^2\delta({\bf
k}-{\bf k'}),
\ee
with all other commutators vanishing. The Feynman
propagator can be now obtained by taking the time-ordered product
of gauge fields, i.e. (recall that $A_{\nu}(x') \equiv A_{\nu'}$),
\be
G_{\mu \nu'}= {i} \,\langle 0 \vert
T\,A_{\mu}(x)\,A_{\nu}(x') \vert 0\rangle\,.
\ee
With the notation of Ref. \cite{PRSLA-A354-79}, one obtains
from Eqs. (4.9)--(4.12) and (4.18)
\be
G_{\mu \nu'}=\left(\begin{array}{cc}
G_{ab'} & 0 \\
0 & G_{ij'} \\
\end{array}\right) ,
\ee
where
\be
G_{ab'}=-\frac{p_a p_{b'}}{\nabla^2} G^{(N)}(x,x')+
\frac{\nabla_a
\nabla_{b'}}{\nabla^2}G^{(D)}(x,x')\;,
\label{Gab}
\ee
\be
G_{ij'}=\delta_{ij}\,G^{(D)}(x,x')\;.
\label{Gij}
\ee

If one follows instead a differential equation approach, one
can verify that the vanishing of off-diagonal blocks in Eq. (4.19)
is also obtainable by finding the kernel of the operator
$$
{\partial^{2}\over \partial \xi^{2}}
+{1\over \xi}{\partial \over \partial \xi}
+\left({\nu^{2}\over \xi^{2}}-k^{2}\right)
$$
and of the operator matrix
\begin{widetext}
\be
M \equiv \left(\begin{array}{cc}
{\partial^{2}\over \partial \xi^{2}}
-{1\over \xi}{\partial \over \partial \xi}
+{\nu^{2}\over \xi^{2}}-k^{2} & -{2i\nu \over \xi} \\
-{2i\nu \over \xi^{3}} &
{\partial^{2}\over \partial \xi^{2}}
+{1\over \xi}{\partial \over \partial \xi}
+{(\nu^{2}-1)\over \xi^{2}}-k^{2} \\
\end{array}\right),
\ee
\end{widetext}
when the boundary conditions (4.2) are imposed. On setting
$\nu \equiv i \mu, \mu \in {\bf R}$, one finds no real roots of the
resulting equations, which involve modified Bessel functions
$I_{\rho}, K_{\rho}$ with $\rho=\mu-1,\mu,\mu+1$. For example, no
real roots exist of the equation
\be
{I_{\rho}(k \xi_{1})\over I_{\rho}(k \xi_{2})}
-{K_{\rho}(k \xi_{1})\over K_{\rho}(k \xi_{2})}=0, \;
\rho=\mu-1,\mu,\mu+1,
\ee
or of the equation
\begin{widetext}
\be
{[I_{\mu}(k \xi_{1})+k I_{\mu}'(k \xi_{1})]\over
[I_{\mu}(k \xi_{2})+k I_{\mu}'(k \xi_{2})]}
-{[K_{\mu}(k \xi_{1})+k K_{\mu}'(k \xi_{1})]\over
[K_{\mu}(k \xi_{2})+k K_{\mu}'(k \xi_{2})]}=0.
\ee
\end{widetext}

Hereafter,
$\nabla^2 \equiv \nabla_i \nabla^i$. The action of the operator
$1/\nabla^2$ in Eq. (4.20) is easily defined, since we shall
require it to act only on functions that have Fourier integral
representation. The ghost Green function is defined by
\be
G(x,x') \equiv {i} \,\langle 0 \vert T\,\chi(x)\,\psi(x') \vert
0\rangle ,
\ee
and is required to obey homogeneous Dirichlet conditions
as we said before, i.e.,
\be
G(x,x')=G^{(D)}(x,x').
\ee

\section{Ward identities}

We now verify that the following Ward identities hold:
\be
G^{\mu}_{\;\,\nu';\mu}+G_{;\nu'}=0,
\label{ward1}
\ee
\be
G^{\mu\nu'}_{\;\;\;;\nu'}+G^{;\mu}=0.
\label{ward2}
\ee
To prove these identities, use is made of the following properties:

\noindent 1) The order of covariant derivatives $\nabla_{\mu}$
can be freely interchanged because the metric is flat, i.e.,
\be
\nabla_{\mu}\nabla_{\nu}=\nabla_{\nu} \nabla_{\mu}.
\ee
\noindent 2) The identity holds
\be
\nabla_a \nabla^a G^{(D/N)}(x,x')=-\nabla^2
G^{(D/N)}(x,x')\;\;\;\;{\rm for}\;x\neq x'
\ee
which easily follows from the Klein--Gordon equation.

\noindent 3) Translation invariance in the $(x,y)$ directions
implies \be \nabla_i G^{(D/N)}(x,x')=-\nabla_{i'} G^{(D/N)}(x,x').
\ee

\noindent 4) since $\epsilon_{ab}$ is antisymmetric and
covariantly constant, $\nabla_a \epsilon_{bc}=0$, it follows that
\be
\nabla^a p_a=\nabla^a
\epsilon_{ab}\nabla^b=\epsilon_{ab}\nabla^a \nabla^b=0.
\ee
By using the above ingredients,
we can easily prove Eq. (\ref{ward1}). Take
first $\nu'=b'$
$$
G^{a}_{\;b';a}+G_{;b'}=\left(\nabla^a
\frac{\nabla_a
\nabla_{b'}}{\nabla^2}+\nabla_{b'}\right)G^{(D)}(x,x')
$$
\be=
\left( -\nabla_{b'}+\nabla_{b'}\right)G^{(D)}(x,x')=0.
\ee
For $\nu'=j'$, we get
\be
G^{i}_{\;j';i}+G_{;j'}=\left(\nabla_j+\nabla_{j'}\right)G^{(D)}(x,x')=0.
\ee
By following analogous steps one proves also Eq. (\ref{ward2}).

\section{Energy-momentum tensors}

Since in what follows we always consider pairs of space-time
points $(x,x')$ with space-like separations, we do not have to
worry about operator ordering, and as a result we can replace in
all formulas the Hadamard function by twice the
imaginary part of the Feynmam propagator.
The Maxwell energy-momentum tensor $T^{\mu \nu}_A$ reads as
\be
T^{\mu
\nu}_{A}=F^{\mu}_{\;\,\beta}F^{\nu \beta}-\frac{1}{4}g^{\mu
\nu}F_{\alpha \beta}F^{\alpha \beta}.
\ee
The gauge and ghost parts of the energy-momentum tensor are
\be
T^{\mu \nu}_{\rm gauge}
=-A^{\alpha\;\;\;\mu}_{\;;\alpha}A^{\nu}
-A^{\alpha\;\;\;\nu}_{\;;\alpha}A^{\mu}+
[A^{\alpha}_{\;\,;\alpha
\beta}A^{\beta}+\frac{1}{2}(A^{\alpha}_{\;\,;\alpha})^2]g^{\mu\nu},
\ee
\be
T^{\mu \nu}_{\rm
ghost}=-\chi^{;\mu}\psi^{;\nu}-\chi^{;\nu}\psi^{;\mu}+g^{\mu
\nu}\chi_{;\alpha}\psi^{;\alpha}.
\ee
By adopting the point-split regularization we define
\be
\langle 0 \vert T^{\mu \nu}_A\vert
0 \rangle \equiv \lim_{x' \rightarrow x}{\cal T}^{\mu \nu}_A(x,x'),
\label{lim1}
\ee
\be
\langle 0 \vert T^{\mu \nu}_{\rm
gauge}\vert 0 \rangle \equiv \lim_{x' \rightarrow x} {\cal T}^{\mu
\nu}_{\rm gauge}(x,x'),
\label{lim2}
\ee
\be
\langle 0 \vert T^{\mu
\nu}_{\rm ghost}\vert 0 \rangle \equiv \lim_{x' \rightarrow x} {\cal
T}^{\mu \nu}_{\rm ghost}(x,x'),
\label{lim3}
\ee
where, on denoting by $g_{\nu}^{\; \mu'}$ the parallel
displacement bivector \cite{Synge1960},
$$
{\cal T}^{\mu \nu}_A(x,x')=\frac{1}{2}g^{\alpha
\beta}(g^{\mu \tau}g^{\nu \rho}
-\frac{1}{4}g^{\mu \nu}g^{\rho\tau})
$$
\be
\times \langle 0 \vert\left\{F_{\tau
\alpha}g^{\rho'}_{\rho} g^{\beta'}_{\beta} F_{\rho' \beta'}
+F_{\rho \beta}g^{\tau'}_{\tau} g^{\alpha'}_{\alpha} F_{\tau'
\alpha'} \right\}\vert 0 \rangle,
\ee
$$
{\cal T}^{\mu \nu}_{\rm gauge}(x,x')=\frac{1}{2} \langle 0
\vert\left\{-A^{\alpha\;\;\;\mu}_{\;;\alpha}
g^{\nu}_{\nu'}A^{\nu'}-\right. A^{\nu}
g_{\mu'}^{\mu}A^{\alpha'\;\;\;\mu'}_{\;;\alpha'}
$$
$$
\left.-A^{\alpha\;\;\;\nu}_{\;;\alpha} g^{\mu}_{\mu'}A^{\mu'}-
A^{\mu} g_{\nu'}^{\nu}A^{\alpha'\;\;\;\nu'}_{\;;\alpha'}+g^{\mu
\nu} [A^{\alpha}_{\;\,;\alpha \beta}
g^{\beta}_{\beta'}A^{\beta'} \right.
$$
\be
\left. + A^{\beta}
g^{\beta'}_{\beta}A^{\alpha'}_{\;\,;\alpha'
\beta'}+A^{\alpha}_{\;\,;\alpha}A^{\alpha'}_{\;\,;\alpha'}]\right\}\vert
0\rangle\ \label{calTgauge},
\ee
and
$$
{\cal T}^{\mu \nu}_{\rm
ghost}(x,x')  =\frac{1}{2} \langle 0 \vert\left\{
-\chi^{;\mu}g^{\nu}_{\nu'}\psi^{;\nu'}
-g^{\mu}_{\mu'}\chi^{;\mu'}\psi^{;\nu}\right.
$$
$$\left.-\chi^{;\nu} g^{\mu}_{\mu'}\psi^{;\mu'}
-g^{\nu}_{\nu'}\chi^{;\nu'}\psi^{;\mu}\right.
$$
\be
\left.+g^{\mu\nu}
(\chi_{;\alpha}g^{\alpha}_{\alpha'}\psi^{;\alpha'}
+g_{\alpha}^{\alpha'}\chi_{;\alpha'}\psi^{;\alpha})
\right\} \vert 0 \rangle
\label{calTghost}.
\ee
Note that ${\cal
T}^{\mu \nu}_{\rm Maxwell}(x,x')$, ${\cal T}^{\mu \nu}_{\rm
gauge}(x,x')$ and ${\cal T}^{\mu \nu}_{\rm ghost}(x,x')$ all
transform as tensors at $x$ and as scalars at $x'$, and therefore
Eqs. (\ref{lim1}-{\ref{lim3}}) are well defined.  By using the Ward
identities Eq. (\ref{ward1}) and Eq. (\ref{ward2}) it is easy to
prove that
\be
{\cal T}^{\mu \nu}_{\rm gauge}(x,x')+{\cal T}^{\mu
\nu}_{\rm ghost}(x,x')=0.
\label{sumzero}
\ee
Indeed, upon
expressing ${\cal T}^{\mu \nu}_{\rm gauge}(x,x')$ and ${\cal
T}^{\mu \nu}_{\rm ghost}(x,x')$ in terms, respectively, of the
photon and ghost propagators $G_{\mu \nu'}$ and $G$,  one can show
that the l.h.s of Eq. (\ref{sumzero}) is equal to
\begin{widetext}
\be
i \left[ g^{(\mu|}_{\mu'}(G^{\alpha
\mu'}_{\;\;\;;\alpha}+G^{;\mu'})^{;|\nu)}+ g^{(\mu}_{\mu'}(G^{\nu)
\alpha'}_{\;\;\;;\alpha'}+G^{;\nu)})^{;\mu'}\right]-\frac{1}{2}g^{\mu
\nu}\left[g^{\beta}_{\beta'}(G^{\alpha
\beta'}_{\;\;\;;\alpha}+G^{;\beta'})_{;\beta'}
+g^{\beta'}_{\beta}(G^{\beta
\alpha'}_{\;\;\;;\alpha'}+G^{;\beta})_{;\beta'}+G^{\alpha
\alpha'}_{\;\;\;;\alpha \alpha'}\right]
\label{vanish},
\ee
\end{widetext}
where indices enclosed within round brackets are symmetrized ( for
example, $A^{(\mu| \lambda \rho| \nu)}:=(A^{\mu \lambda \rho
\nu}+A^{\nu \lambda \rho \mu})/2$). The four terms between the
round brackets coincide with the l.h.s of either Eq. (\ref{ward1})
or Eq. (\ref{ward2}), and hence vanish. As for the last term
between the second pair of square brackets of the above
expression, Eq. (\ref{ward2}) allows to write it as \be G^{\alpha
\alpha'}_{\;\;\;;\alpha \alpha'}=- \cstok{\ }\,G \ee and
therefore, since $x \neq x'$, it vanishes because of Eq.
(\ref{eom}).

\section{The Maxwell energy-momentum tensor}

According to Eq. (\ref{sumzero}), the gauge and ghost
contributions to the total energy-momentum tensor cancel each other,
and therefore the Maxwell energy-momentum tensor is the only contribution
that needs to be considered. Its evaluation is made easy by making
use of the representation of the scalar Green functions $G^{(D)}$
and $G^{(N)}$ given in Eq. (\ref{split}). Using it, we can express
the photon propagator $G_{\mu \nu'}$ as the sum of two terms
\be
G_{\mu \nu'}=G^{(0)}_{\mu \nu'}+\tilde{G}_{\mu
\nu'},.
\label{splitG}
\ee
The first term coincides with the
covariant photon propagator of the entire Minkowski manifold,
transformed to Rindler coordinates
\be
G^{(0)}_{\mu \nu'}=g_{\mu
\nu}g^{\nu}_{\nu'}G_0,
\ee
$g^{\mu}_{\nu'}$ denoting the bivector
of parallel displacement from $x'$ to $x$ along any arc connecting
$x'$ to $x$ \footnote{Indeed, since the Rindler metric is flat,
parallel displacement is independent of the choice of the arc
connecting $x$ to $x'$.} (see Appendix B). As for the second
term $\tilde{G}_{\mu \nu'}$, it is equal to
\be
\tilde{G}_{\mu \nu'}=\left(\begin{array}{cc}
\tilde{G}_{ab'} & 0 \\
0 & \tilde{G}_{ij'} \\
\end{array}\right),
\ee
where
\be
\tilde{G}_{ab'}=-\frac{p_a p_{b'}}{\nabla^2} \tilde{G}^{(N)}(x,x')+
\frac{\nabla_a
\nabla_{b'}}{\nabla^2}\tilde{G}^{(D)}(x,x')\;,
\label{Gab}
\ee
\be
\tilde{G}_{ij'}=\delta_{ij}\,\tilde{G}^{(D)}(x,x')\;.
\label{Gij}
\ee
The important thing to notice is that the singularities of the
photon propagator are all included in the $G^{(0)}_{\mu \nu'}$
piece, while $\tilde{G}_{\mu \nu'}$ is perfectly regular in the
coincidence limit $x' \rightarrow x$.  The Maxwell tensor admits a
representation analogous to Eq. (\ref{splitG}), i.e.,
\be
{\cal T}^{\mu
\nu}_A(x,x')={\cal T}^{(0)\mu \nu}_A(x,x')+\widetilde{{\cal T}}^{\mu
\nu}_A(x,x').
\ee
Here, ${\cal T}^{(0)\mu \nu}_A(x,x')$ is the
contribution arising from $G^{(0)}_{\mu \nu'}$, while
$\widetilde{{\cal T}}^{\mu \nu}_A(x,x')$ is the contribution involving
${\tilde{G}}_{\mu \nu'}$. The quantity ${\cal T}^{(0)\mu
\nu}_A(x,x')$ coincides with the point-split expression for the
Maxwell tensor in Minkowski spacetime, transformed to Rindler
coordinates, and it diverges in the limit $x'\rightarrow x$. Being
independent of the plates' separation $a$, we shall simply
disregard it. On the contrary, the expression $\widetilde{{\cal
T}}^{\mu \nu}_A(x,x')$ is perfectly well defined in the
coincidence limit. In this limit, its explicit expression is
\cite{PRSLA-A354-79}
\be
\langle 0 \vert \widetilde{T}^{\mu}_{A \nu}\vert
0 \rangle= {\rm diag}\,
(-\gamma+\delta,\gamma,\gamma,-\gamma-\delta),
\label{treg}
\ee
where
\be
\delta=\left.\frac{i  }{2}\left(
\frac{\xi_1^2}{\xi \xi'}\frac{\partial^2}{\partial t
\partial t'}+\frac{\partial^2}{\partial \xi
\partial
\xi'}\right)\right\vert_{\xi'=\xi}\!\!\!\!
(\tilde{G}^{(N)}+\tilde{G}^{(D)}),
\label{delta0}
\ee
\be
\gamma=\frac{i}{2}\nabla^2(\tilde{G}^{(N)}
+\tilde{G}^{(D)}).
\label{gamma0}
\ee
It is clear from the above formulas that $\langle 0 \vert
\tilde{T}^{\mu}_{A \nu}\vert 0 \rangle$ is traceless, i.e.,
\be
\langle
0 \vert \tilde{T}^{\mu}_{A \mu}\vert 0 \rangle =0.
\label{tless}
\ee
It can also be verified that $\gamma$ and $\delta$ satisfy the relation
\be
-\gamma+\delta=-\frac{d}{d\xi}[\xi (\gamma+\delta)],
\label{cons}
\ee
which represents the
condition for ${\widetilde T}^{\mu \nu}_A$
to be covariantly conserved.
%As soon as
%either one of the quantities $\gamma$ and  $\delta$ is known, the
%other can be obtained by solving Eq. (\ref{cons}). If we take
%$\gamma$ as the independent quantity, by solving Eq. (\ref{cons})
%we obtain
%\be
%\delta(\xi)=-\frac{1}{\xi^2} \int dx\,x^2\,
%\frac{d\gamma}{d x}(x)+\frac{1}{a^4}\,\frac{\xi_1^2}{\xi^2}\,
%F\left(\frac{a}{\xi_1}\right),
%\ee
%where $F(a/\xi_1)$ is a
%function of $a/\xi_1$ to be determined.
The expressions for
$\gamma$ and $\delta$ can be obtained by inserting Eq. (\ref{tildeG})
into Eqs. (\ref{delta0}) and (\ref{gamma0}), i.e.,
\be
\left.\delta=\frac{i a }{2  }\!\int\!\! \frac{d \omega}{2 \pi}
\int \!\!\!\frac{d^2{\bf k}}{(2 \pi)^2}  \left( \frac{\xi_1^2
\omega^2}{\xi \xi'}+\frac{\partial^2}{\partial \xi
\partial \xi'}\right)  \tilde{\psi} (\xi,\xi'|i \xi_1
\omega,k)\right\vert_{\xi'=\xi}\!\!,\label{deltafin} \ee \be
\gamma=- i \frac{a}{2}\!\int\!\! \frac{d \omega}{2 \pi} \int
\!\!\frac{d^2{\bf k}}{(2 \pi)^2} k^2  \tilde{\psi} (\xi,\xi|i
\xi_1 \omega,k),\label{gammafin} \ee where we have defined \be
\tilde{\psi} (\xi,\xi'|i \xi_1 \omega,k) \equiv
\frac{\xi_1}{a}(\tilde{\chi}^{(D)} +\tilde{\chi}^{(N)})
(\xi,\xi'|i \xi_1 \omega,k) \ee It is convenient to rotate the
contour of the $\omega$-integration away from the singularities to
the positive imaginary axis. Since ${\tilde \psi}$ is an even
function of $\omega$, and is invariant under rotations in the
$(x,y)$ plane, upon setting $\omega \equiv i \eta$ and then
performing the change of variables $a k \equiv q \sqrt{1-s^2}$,
$\eta \equiv s q/a$, the integrals for $\delta$ and $\gamma$
become \be \delta= \frac{1}{4 \pi^2 a^4}\left.
\!\!\int_0^{\infty}\!\! \!\!\!d q \,q^2
 \!\!\int_0^1
\!\! ds \left(  {s}^2 q^2\frac{{\hat \xi}_1^2}{{\hat \xi} {\hat
\xi'}}-\frac{\partial^2}{\partial {\hat \xi}
\partial
{\hat \xi'}}\right)  \tilde{\psi} (\xi,\xi')\right\vert_{\xi'=\xi}
\!\!,
\label{delta}
\ee
\be
\gamma=
\frac{1}{4 \pi^2 a^4}\int_0^{\infty}\!\! \!\!\!d q \,q^4
 \!\!\int_0^1
\!\! ds (1-s^2) \tilde{\psi} (\xi,\xi), \label{gamma} \ee where we
have set ${\hat \xi} \equiv \xi/a$, ${\hat \xi}_i \equiv
\xi_i/a,\,\,i=1,2$. The weak-field limit is obtained by taking
$\hat{\xi}_1 \rightarrow \infty$ in the previous formulas, for
fixed $s$ and $q$. By using the large-order uniform asymptotic
expansions of the modified Bessel functions, quoted in Appendix A,
we have obtained the asymptotic expansion for $\tilde{\psi}$, to
second order in ${\rm g} a$ (hereafter ${\hat z} \equiv z/a, {\hat
z}' \equiv z'/a$): \be \tilde{\psi}(\hat{z},\hat{z}') \sim
\tilde{\psi}^{(0)}+{\rm g} a \,\tilde{\psi}^{(1)}+({\rm g} a)^2
\,\tilde{\psi}^{(2)}+{\rm O}(({\rm g}a)^3), \ee where \be
\tilde{\psi}^{(0)}=\frac{e^{q(\hat{z}-\hat{z}')}+e^{q(\hat{z}'-\hat{z})}}{q(e^{2
q}-1)}\,\ee and
\begin{widetext}
\be \tilde{\psi}^{(1)}
=\frac{\{e^{q(\hat{z}+\hat{z}')}-e^{q(2-\hat{z}-\hat{z}')}
-2q(\hat{z}+\hat{z}') \cosh[q(\hat{z}-\hat{z}')]
\}(1-s^2)-2s^2q^2(\hat{z}^2-\hat{z}'^2) \sinh[q(\hat{z}-\hat{z}')]
}{2 q^2(e^{2q}-1)}+ \frac{\cosh[q(\hat{z}-\hat{z}')] }{2
\sinh^2(q)}s^2. \ee The expression for $\tilde{\psi}^{(2)}$ is
exceedingly lengthy  and will not be reported here.
\end{widetext}
Evaluation of the integrals then gives the result
\begin{widetext}
$$ \delta \sim \frac{\pi^2}{360 a^4} +\frac{\rm g}{a^3}
\left(\frac{\pi^2}{450}(1-2{\hat z}) +\frac{\pi}{60}
\frac{\cos(\pi {\hat z})}{\sin^{3}(\pi{\hat z})} \right)+\frac{\rm
g^2}{a^2}\left[\frac{\pi^2 (1-104 \hat{z}+160 \hat{z}^2) -160}{
25200}-\frac{\pi^2 \hat{z}(\hat{z}-1)-8}{420 \sin^2(\pi \hat{z})}
\right.$$ \be\left.-\frac{\pi (20 \hat{z}-3)\cos(\pi \hat{z})}{840
\sin^3(\pi \hat{z})}-\frac{\pi^2 \hat{z}(1-\hat{z})}{280
\sin^4(\pi \hat{z})}\right] +{\rm O}({\rm g}^{3}),
\ee
\end{widetext}
\begin{widetext}
$$\gamma \sim
\frac{\pi^2}{720 a^4}   +\frac{\rm g}{a^3} \left(
\frac{\pi^2}{1800}(1-2\hat{z})-\frac{\pi}{60} \frac{\cos(\pi
\hat{z})}{\sin ^3 (\pi \hat{z})} \right) +\frac{\rm g^2}{a^2}
\left[ \frac{\pi^2 (44 \hat{z}^2-16
\hat{z}-9)-100}{50400}+\frac{\pi^2 \hat{z}(\hat{z}-1)-1}{420
\sin^2(\pi \hat{z})}    \right.$$ \be \left.+\frac{\pi (20
\hat{z}-3)\cos(\pi \hat{z})}{840 \sin^3(\pi \hat{z})}+\frac{\pi^2
\hat{z}(1-\hat{z})}{280 \sin^4(\pi \hat{z})}
 \right] + {\rm O}({\rm g}^{3}).
\ee
\end{widetext}
It can be verified that the above expressions for $\gamma$ and $\delta$
satisfy the fundamental conservation condition Eq. (\ref{cons}).
Moreover, on inserting these values into Eq. (\ref{treg}) we obtain
\begin{widetext}
$$ \langle 0 \vert \widetilde{T}^{\;\;\,t}_{A t}\vert 0
\rangle \sim \frac{\pi^2}{720a^4} +\frac{\rm g}{a^3}
\left(\frac{\pi^2}{600}(1-2{\hat z})+\frac{\pi}{30} \frac{\cos(\pi
{\hat z})}{\sin^{3}(\pi {\hat z})} \right)+\frac{\rm
g^2}{a^2}\left[-\frac{11}{2520}+\frac{\pi^2}{50400}(11-192
\hat{z}+276 \hat{z}^2)\right.$$ \be\left. + \frac{9-2 \pi^2
\hat{z}(\hat{z}-1)}{420 \sin^2(\pi \hat{z})}-\frac{\pi (20
\hat{z}-3)\cos(\pi \hat{z})}{420 \sin^3(\pi \hat{z})}-\frac{\pi^2
\hat{z}(1-\hat{z})}{140 \sin^4(\pi \hat{z}) }\right] +{\rm O}({\rm
g}^{3}), \label{Ttt} \ee \be \langle 0 \vert
\widetilde{T}^{\;\,\,z}_ {Az}\vert 0 \rangle
\sim-\frac{\pi^2}{240a^4}-\frac{\rm g}{a^3}
\frac{\pi^2}{360}(1-2{\hat z})+\frac{\rm
g^2}{a^2}\left\{\frac{1}{120}+\frac{\pi^2}{7200}[1+4 \hat{z}(8-13
\hat{z})]-\frac{1}{60 \sin^2(\pi \hat{z})}\right\} +{\rm O}({\rm
g}^{3}), \ee
\end{widetext}
while of course
\be
\langle 0 \vert
\widetilde{T}^{\;\;\,x}_{A x}\vert 0 \rangle= \langle 0 \vert
\widetilde{T}^{\;\;\,y}_{A y}\vert 0 \rangle=\gamma\,.
\label{Txx}
\ee
We note that the quantities $\gamma$ and $\delta$
both diverge as $z$ approaches the locations of the plates at
$z=0$ and $z=a$. In particular, for ${\hat z}\rightarrow 0$, from
Eqs. (\ref{Ttt}) and (\ref{Txx}) we find
\be
\langle 0 \vert
\widetilde{T}^{\;\;\,t}_{A t}\vert 0 \rangle\sim \frac{{\rm g}}{30
\pi^2 z^3}+O({z}^{-2}),
\ee
\be
\langle 0 \vert
\widetilde{T}^{\;\;\,z}_{A z}\vert 0 \rangle\sim -\frac{{\rm
g^2}}{60 \pi^2 z^2}+O({z}^{-1}),
\ee
\be
\langle 0 \vert
\widetilde{T}^{\;\;\,x}_{A x}\vert 0\rangle= \langle 0 \vert
\widetilde{T}^{\;\;\,y}_{A y}\vert 0\rangle \sim -\frac{{\rm
g}}{60 \pi^2 z^3}+O({z}^{-2}).
\ee
These behaviors are in full
agreement with the results derived in Ref. \cite{PRSLA-A354-79},
for the case of a single mirror \footnote{When comparing our
formulae with those of \cite{PRSLA-A354-79}, our $\xi_1$
corresponds with the $a$ of \cite{PRSLA-A354-79}}. The valuable
work in Ref. \cite{PHRVA-D66-085023}, devoted to the scalar and
electromagnetic Casimir effects in the Fulling--Rindler vacuum,
can also be shown to agree with our energy-momentum formulas.

\section{Concluding remarks}

Our analysis has made it possible to put on completely firm ground
the set of formulas for the vacuum expectation value of the
regularized and renormalized energy-momentum tensor for an
electromagnetic Casimir apparatus in a weak gravitational field.
In particular, the term of first order in ${\rm g}$ in
Eq. (\ref{Ttt}) corrects an unfortunate mistake in
Eq. (4.4) of Ref. \cite{PHRVA-D74-085011} (see Ref. \cite{erratum}).
Using our original Eqs. (\ref{delta}) and
(\ref{gamma}) we have been able to evaluate second-order
corrections (with respect to the expansion parameter ${\rm g}
a/c^2$) to $\langle T_{\mu \nu} \rangle$, which represent one new
result of the present paper. The physical interpretation that can
be attributed to these corrections is doubtful, in view of the
divergencies they exhibit on approaching the plates. The existence
of these divergencies is well known in the literature \cite{PHRVA-D20-3063},
and it is usually attributed to the pathological character of
perfect-conductor boundary conditions. Indeed, divergencies arise
already in first order corrections to some components of $T_{\mu
\nu}$, but they constitute somewhat less of a problem, because
while on the one hand no divergence is found in $T_{zz}$, which
provides the Casimir pressure, the nonintegrable divergencies in
$T_{tt}$ are of such a nature that one can still obtain a finite
value for the total mass-energy of the Casimir apparatus (per unit
area of the plates), by taking the principal-value integral of
$T^t_t$ over the volume of the cavity \cite{PHRVA-D74-085011}.
Neither of these fortunate circumstances occurs at second order,
since on the one hand $T_{zz}$ is now found to diverge on approaching
the plates, so that no definite meaning can be given to the
gravitational correction to the Casimir pressure, and on the other
hand the divergencies in $T^t_t$ are such that the resulting
correction to the total mass-energy of the cavity is infinite,
even on taking the principal-value integral of $T^t_t$.

The years to come will hopefully tell us whether the push
predicted and confirmed by theory is amenable to experimental
verification \cite{PHLTA-A297-328}. It also remains to be seen
whether the experience gained in the detailed evaluation of the
energy-momentum tensor in Ref. \cite{PHRVA-D74-085011} and in the
present paper can be used to obtain a better understanding of the
intriguing relation between Casimir effect and Hawking radiation
found in Ref. \cite{CQGRD-18-2097}.

\acknowledgments We are grateful to E. Calloni for previous
collaboration, and to S. Fulling, K. Milton and A. Saharian for
scientific correspondence and conversations that prompted us to
double-check our early work on the Maxwell energy-momentum tensor.
G. Esposito is grateful to the Dipartimento di Scienze Fisiche of
Federico II University for its hospitality and support. The work
of L. Rosa has been partially supported by PRIN {\it Fisica
Astroparticellare}.

\appendix

\section{Asymptotic formulas}

For large orders $\nu$, the modified Bessel functions
$I_{\nu}(\nu w)$, $K_{\nu}(\nu w)$ and their first derivatives
admit the following asymptotic expansions, which hold uniformly
with respect to $w$ in the half-plane $|{\rm arg}w | \leq
{\pi \over 2}-\varepsilon$, for $\varepsilon$ in the open interval
$\left ]0,{\pi \over 2} \right [$ \cite{PTRSA-A247-328}, \cite{abram}:
\be
I_{\nu}(\nu w)\sim
\frac{1}{\sqrt{2 \pi \nu}}\frac{e^{\nu
\rho}}{(1+w^2)^{1/4}}\left\{
1+\sum_{k=1}^{\infty}\frac{u_k(t)}{\nu^k}\right\},
\ee
\be
K_{\nu}(\nu w)\sim  \sqrt{\frac{\pi}{2 \nu}}\frac{e^{-\nu
\rho}}{(1+w^2)^{1/4}}\left\{
1+\sum_{k=1}^{\infty}(-1)^k\frac{u_k(t)}{\nu^k}\right\},
\ee
\be
I'_{\nu}(\nu w)\sim \frac{1}{\sqrt{2 \pi
\nu}}\frac{(1+w^2)^{1/4}}{w}{e^{\nu \rho}}\left\{
1+\sum_{k=1}^{\infty}\frac{v_k(t)}{\nu^k}\right\},
\ee
\be
K'_{\nu}(\nu w)\sim  -\sqrt{\frac{\pi}{2
\nu}}\frac{(1+w^2)^{1/4}}{w} {e^{-\nu \rho}} \left\{
1+\sum_{k=1}^{\infty}(-1)^k\frac{v_k(t)}{\nu^k}\right\},
\ee
where
\be
t \equiv 1/\sqrt{1+w^2}
\ee
\be
\rho \equiv \sqrt{1+w^2}+\log
\frac{w}{1+\sqrt{1+w^2}},
\ee
and, for $k =0,1,2$, one has
\begin{eqnarray}
v_0 &=& 1, \\
v_1 &=& (-9 t+7 t^3)/24. \\
v_2 &=& (-135 t^2+594 t^4-455 t^6)/1152,
\end{eqnarray}
\begin{eqnarray}
u_0 &=& 1, \\
u_1 &=& (3 t-5 t^3)/24, \\
u_2 &=& (81 t^2-462 t^4+385 t^6)/1152,
\end{eqnarray}
The generating formulas of these Olver polynomials
\cite{PTRSA-A247-328} are
\be
u_{k+1}(t)={t^{2}\over 2}(1-t^{2}){du_{k}\over dt}
+{1\over 8}\int_{0}^{t}(1-5 \beta^{2})u_{k}(\beta)d\beta,
\ee
\be
v_{k}(t)=u_{k}(t)-t(1-t^{2})
\left({1\over 2}u_{k-1}+t {du_{k-1}\over dt}\right).
\ee

\section{The bivector of parallel displacement}

The bivector of parallel displacement $\delta^{\mu'}_{\nu}$
\cite{Synge1960} in the
Rindler spacetime is easily evaluated by exploiting the
coordinate transformation
$$
\bar{\tau}=\xi\,\sinh
\tau\;,\;\;\;\bar{z}=\xi \cosh{\tau}\;,
$$
\be
\bar{x}=x\;,\;\;\;\;\bar{y}=y,
\ee
where $\tau=t/\xi_1$, that turns
the Rindler metric in Eq. (2.2) into the Minkowski metric
\be
ds^2=-d \bar{\tau}^2+d \bar{x}^2+d \bar{y}^2+d\bar{y}^2.
\ee
In the Minkowski coordinates we obviously have
$\bar{g}^{\mu'}_{\nu}=\delta^{\mu'}_{\nu}$. Therefore
\be
{g}^{\mu'}_{\nu}=\bar{g}^{\rho'}_{\sigma} \,\frac{\partial
\bar{x}^{\sigma}}{\partial x^{\nu}}\, \frac{\partial
{x'}^{\mu}}{\partial \bar{x}'^{\rho}}=\frac{\partial
\bar{x}^{\rho}}{\partial x^{\nu}}\, \frac{\partial
{x}'^{\mu}}{\partial \bar{x}'^{\rho}}.
\ee
We then obtain
\be
g^{\mu'}_{\nu}=\left(\begin{array}{cc}
g^{a'}_b & 0 \\
0 & \delta^{i'}_j \\
\end{array}\right),
\ee
where
\begin{widetext}
\be
g^{a'}_{b}=\left(\begin{array}{cc}
g^{\tau'}_{\tau} & g^{\tau'}_{\xi} \\
g^{\xi'}_{\tau} & g^{\xi'}_{\xi} \\
\end{array}\right)=\left(\begin{array}{cc}
\frac{\xi}{\xi'}(\cosh \tau' \cosh \tau-\sinh \tau' \sinh \tau)
& \frac{1}{\xi'}(\cosh \tau' \sinh \tau-\sinh \tau' \cosh \tau) \\
\xi (\cosh \tau' \sinh \tau-\sinh \tau' \cosh \tau)
& \cosh \tau' \cosh \tau-\sinh \tau' \sinh \tau\\
\end{array}\right).
\ee
\end{widetext}
Similarly, one finds
\be
g^{\mu}_{\nu'}=g^{\mu \rho}g_{\nu'
\sigma'}g^{\sigma'}_{\rho}=\left(\begin{array}{cc}
g^{a}_{b'} & 0 \\
0 & \delta^{i}_{j'} \\
\end{array}\right),
\ee
where
\begin{widetext}
\be
g^{a}_{b'}=\left(\begin{array}{cc}
g^{\tau}_{\tau'} & g^{\tau}_{\xi'} \\
g^{\xi}_{\tau'} & g^{\xi}_{\xi'} \\
\end{array}\right)=\left(\begin{array}{cc}
\frac{\xi'}{\xi}(\cosh \tau' \cosh \tau
-\sinh \tau' \sinh \tau) &
-\frac{1}{\xi}(\cosh \tau' \sinh \tau-\sinh \tau' \cosh \tau) \\
-\xi' (\cosh \tau' \sinh \tau-\sinh \tau' \cosh \tau)
&  \cosh \tau' \cosh \tau-\sinh \tau' \sinh \tau\\
\end{array}\right).
\ee
\end{widetext}
It can be checked that
\be
g^{\mu}_{\rho'}g^{\rho'}_{\nu}=\delta^{\mu}_{\nu}.
\ee

\end{document}